\documentclass[a4paper,11pt]{article}
\usepackage{amsmath,amsthm,amssymb}
\usepackage{graphicx,subfigure}
\usepackage{url}

\usepackage{algorithm2e}
\usepackage{listings}
\usepackage{color}

\usepackage{multicol}
\def\twoplot[#1]#2#3#4#5{
\begin{figure}[h]
\begin{multicols}{2}
\begin{center}
    \includegraphics*[#1]{#2}
    \caption{\label{#2} #4}
\end{center}
\begin{center}
    \includegraphics*[#1]{#3}
    \caption{\label{#3} #5}
\end{center}
\end{multicols}
\end{figure}
}

\makeatletter
\def\@citex[#1]#2{\immediate\write\@auxout{\string\citation{#2}}
\def\@citea{}\@cite{\@for\@citeb:=#2\do
{\@citea\def\@citea{; }\@ifundefined
{b@\@citeb}{{\bf ?}\@warning
{Citation `\@citeb' on page \thepage \space undefined}}%
{\csname b@\@citeb\endcsname}}}{#1}}
\makeatother

\usepackage[colorlinks,bookmarksopen,bookmarksnumbered,citecolor=red,urlcolor=red,breaklinks=true]{hyperref}
\hypersetup{pdftitle={Spatial Global Sensitivity Analysis of High Resolution classified topographic data use
in 2D urban flood modelling},
bookmarks=true,
pdftoolbar=true,
pdfmenubar=true,
pdfauthor={M. Abily, N. Bertrand, O. Delestre, P. Gourbesville, C.-M. Duluc},
pdfsubject={Urabn flood, Uncertainties, flood hazard modelling, FullSWOF, sensitivity maps, photogrammetry,
classified data, sensitivity analysis, shallow water equations, Sobol index},
pdfcreator={Delestre},
pdfproducer={Delestre},
pdfkeywords={uncertainty} {flood hazard} {Shallow-Water equation} {Saint-Venant} {global sensitivity analysis} {Sobol index}
{high resolution modelling} {DSM} {surface model} {photogrammetry} {urban flood} {sensitivity maps} {photogrammetry}
 {classified topographic data} {FullSWOF} {Var river} {finite volume} {well-balanced} {hydrostatic reconstruction}}

\title{Spatial Global Sensitivity Analysis of High Resolution classified topographic data use in 2D
urban flood modelling }

\author{{M. Abily}\footnote{Polytech'Nice Sophia \& URE Innovative-CiTy, University of Nice Sophia Antipolis,
France, e-mail : abily@polytech.unice.fr}, {N. Bertrand}\footnote{Institut
de Radioprotection et de S\^uret\'e Nucl\'eaire
(IRSN), PRP-DGE, SCAN, BEHRIG, France, e-mail : nathalie.bertrand@irsn.fr},
{O. Delestre}\footnote{Lab. J.A. Dieudonn\'e UMR7351 CNRS \& EPU Nice
Sophia, University of Nice, France, e-mail : delestre@math.unice.fr},\\
P. Gourbesville\footnote{Polytech'Nice
Sophia \& URE Innovative-CiTy, University of Nice Sophia Antipolis, France},
\; and 
{C.-M. Duluc}\footnote{Institut de Radioprotection et de S\^uret\'e Nucl\'eaire
(IRSN), PRP-DGE, SCAN, BEHRIG}}

\begin{document}
\maketitle

\begin{abstract}
This paper presents a spatial Global Sensitivity Analysis (GSA) approach in a 2D shallow water equations
based High Resolution (HR) flood model. The aim of a spatial GSA is to produce sensitivity maps which
are based on Sobol index estimations. Such an approach allows to rank the effects of uncertain HR
topographic data input parameters on flood model output. The influence of the three following
parameters has been studied: the measurement error, the level of details of above-ground elements
representation and the spatial discretization resolution. To introduce uncertainty, a Probability
Density Function and discrete spatial approach have been applied to generate $2,000$ DEMs. Based
on a 2D urban flood river event modelling, the produced sensitivity maps highlight the major influence
of modeller choices compared to HR measurement errors when HR topographic data are used, and the
spatial variability of the ranking.
\end{abstract}

\section*{Highlights}

\begin{itemize}
 \item Spatial GSA allowed the production of Sobol index maps, enhancing the relative weight of each
 uncertain parameter on the variability of calculated output parameter of interest.
 \item The Sobol index maps illustrate the major influence of the modeller choices, when using the HR
 topographic data in 2D hydraulic models with respect to the influence of HR dataset accuracy.
 \item Added value is for modeller to better understand limits of his model. 
 \item Requirements and limits for this approach are related to subjectivity of choices and to computational cost.
\end{itemize}

\section*{Softwares availability}

\begin{equation*}
 \begin{array}{|c|c|c|}\cline{1-3}
  \text{ {\tiny  Name of softwares}} & \text{{\tiny FullSWOF\_2D}} & \text{{\tiny Prom\'eth\'ee}} \\ \hline
  \text{{\tiny Developers}} & \begin{array}{c}
  \text{{\tiny F. Darboux, O. Delestre,}}\\
  \text{{\tiny C. Laguerre, C. Lucas, M.H. Le}}\end{array} & \text{{\tiny IRSN}}\\ \hline
  \text{{\tiny Contacts}} & \text{{\tiny \url{fullswof.contact@listes.univ-orleans.fr}}} &
  \text{{\tiny \url{Yann.richet@irsn.fr}}}\\ \hline
  \text{{\tiny First year of availibility}} & \text{{\tiny 2011}} & \text{{\tiny 2009}}\\ \hline
  \text{{\tiny Operating system}} & \text{{\tiny Linux, Windows and Mac}} & \text{{\tiny Linux, Windows and Mac}}\\ \hline
  \text{{\tiny Software availibility}} & \text{{\tiny \url{https://sourcesup.renater.fr/}}} & \text{{\tiny \url{http://promethee.irsn.fr}}} \\ \hline
  \text{{\tiny Cost}} & \begin{array}{c}
  \text{{\tiny Free of charge, distributed under}}\\
  \text{{\tiny CeCILL-V2 license (GPL compatible)}}\end{array} & \text{{\tiny Free of charge}}\\ \hline
 \end{array}
\end{equation*}

\section{Introduction}\label{sec:intro}
In hydraulics, deterministic numerical modelling tools based on approximating solutions of the 2D
Shallow Water Equations (SWE) system are commonly used for flood hazard assessment
\cite{Gourbesville15}. This category of tools describes water free surface behavior (mainly
elevation and discharge) according to an engineering conceptualization, aiming to provide to decision
makers information that often consists in a flood map of maximal water depths. As underlined in
\cite{Cunge14}, good practice in hydraulic numerical modelling is for modellers to know in detail the
chain of concepts in the modelling process and to supply to decision makers possible doubts and
deviation between what has been simulated and the reality. Indeed, in considered SWE based models,
sources of uncertainties come from ($i$) hypothesis in the mathematical description of the natural
phenomena, ($ii$) numerical aspects when solving the model, ($iii$) lack of knowledge in input parameters
and ($iv$) natural phenomena inherent randomness. Errors arising from $i$, $ii$ and $iii$ may be considered
as belonging to the category of epistemic uncertainties (that can be reduced {\it e.g.} by improvement
of description, measurement). Errors of type $iv$ are seen as stochastic errors (where randomness is
considered as a part of the natural process, {\it e.g.} in climatic born data) \cite{Walker03}.
At the same time, the combination of the increasing availability of High Resolution (HR) topographic
data and of High Performance Computing (HPC) structures, leads to a growing production of HR flood
models \cite{Abily13b,Erpicum10,Fewtrell11,Hunter08,Meesuk15}. For non-practitioner, the level of accuracy of HR
topographic data might be erroneously interpreted as the level of accuracy of the HR flood models, disregarding uncertainty
inherent to this type of data use, not without standing the fact that other types of above mentioned
errors occur in hydraulic modelling. 

\subsection{High Resolution topographic data and associated errors}\label{subsec:High Resolution
topographic data and associated errors}

Topographic data is a major input for flood models, especially for complex environment such as urban
and industrial areas, where a detailed topography helps for a better description of the physical
properties of the modelled system \cite{Abily13b,Djordjevic13,Gourbesville15}.
In the case of an urban or industrial environment, a topographic dataset is considered to be of HR
when it allows to include in the topographic information the elevation of infra-metric elements
\cite{LeBris13}. These infra-metric elements (such as sidewalks, road-curbs, walls, {\it etc.})
are features that influence flow path and overland flow free surface properties. At megacities scale,
HR topographic datasets are getting commonly available at an infra-metric resolution using modern
gathering technologies (such as LiDAR, photogrammetry) through the use of aerial vectors like
unmanned aerial vehicle or specific flight campaign \cite{Chen09,Meesuk15,Musialski13,Nex13,Remondino11}. Moreover, modern urban
reconstruction methods based on features classification carried out by photo-interpretation process,
allow to have high accuracy and highly detailed topographic information \cite{Andres12,Lafarge10,Lafarge11,Mastin09}.
Photo-interpreted HR datasets allow to generate HR DEMs including classes of impervious above ground
features \cite{Abily14b}. Therefore generated HR DEMs can include above ground features elevation information depending on
modeller selection among classes. Based on HR classified topographic datasets, produced HR Digital
Elevation Model (DEM) can have a vertical and horizontal accuracy up to $0.1\;\text{m}$ \cite{Fewtrell11}.
\newline

Even though being of high accuracy, produced HR DEMs are assorted with the same types of errors as
coarser DEMs. Errors are due to limitations in measurement techniques and to operational restrictions.
These errors can be categorized as: ($i$) systematic, due to bias in measurement and processing; ($ii$)
nuggets (or blunder), which are local abnormal value resulting from equipment or user failure, or to
occurrence of abnormal phenomena in the gathering process ({\it e.g.} birds passing between the ground
and the measurement device) or ($iii$) random variations, due to measurement/operation inherent limits
(see \cite{Fisher06,Wechsler07}). Moreover, the amount of data that composes a HR classified
topographic dataset is massive. Consequently, to handle the HR dataset and to avoid prohibitive
computational time, hydraulic modellers make choices to integrate this type of data in the hydraulic
model, possibly decreasing HR DEM quality and introducing uncertainty \cite{Tsubaki13,Abily15}. As recalled
in the literature \cite{Dottori13,Tsubaki13}, in HR flood models, effects of uncertainties related to HR topographic
data use on simulated flow is not yet quantitatively understood. 

\subsection{Uncertainty and Sensitivity Analysis}\label{subsec:Uncertainty and Sensitivity Analysis}

To evaluate uncertainty in deterministic models, Uncertainty Analysis (UA) and Sensitivity Analysis (SA)
have started to be used \cite{Saltelli00} and \cite{Saltelli08} and become broadly applied for a wild range
of environmental modelling problems \cite{Refsgaard07,Uusitalo15}. UA consists
in the propagation of uncertainty sources through model, and then focuses on the quantification
of uncertainties in model output allowing robustness to be checked \cite{SaintGeours12}. SA aims
to study how uncertainty in a model output can be linked and allocated proportionally to the
contribution of each input uncertainties. Both UA and SA are essential to analyze complex systems
\cite{Helton06,SaintGeours14}, as study of uncertainties related to input
parameters is of prime interest for applied practitioners willing to decrease uncertainties in their
models results \cite{Iooss11}.
\newline

In 1D and 2D flood modelling studies, approaches based on sampling based methods are becoming used in
practical applications for UA. For SA, depending on applications and objectives, different categories
of variance based approaches have been recently applied in flood modelling studies (mainly in 1D)
such as Local Sensitivity Analysis (LSA) \cite{Delenne12} or more recently, a Global
Sensitivity Analysis (GSA) based on a screening method has been implemented in 2D flood modelling application
\cite{Willis14}.

\subsubsection*{Local Sensitivity Analysis}

LSA focuses on fixed point in the space of the input and aims to address model behavior near parameters
nominal value to safely assume local linear dependences on the parameter. LSA can use either a
differentiation or a continuous approach \cite{Delenne12}. LSA based on differentiation
approach performs simulations with slight differences in a given input parameter and computes
the difference in the results variation, with respect to the parameter variation. LSA based on
continuous approach differentiates directly the equations of the model, creating sensitivity equation
\cite{Delenne12}. The advantages of LSA approaches are that they are not resource demanding in
terms of computational cost, drawback being that the space of input is locally explored assuming
linear effects only. Linear effects means that given change in an input parameter introduces a
proportional change in model output, in opposition to nonlinear effects. LSA approaches perform
reasonably well with SWE system even if nonlinear effects occur punctually (see \cite{Delenne12}).
Nonetheless, important nonlinear effects in model output might arise when parameters are interacting and
when solution becomes discontinuous. LSA consequently becomes not suited \cite{Delenne12,Guinot07} in
such a context, which is likely to occur in case of 2D SWE based simulation of overland flow. 

\subsubsection*{Global Sensitivity Analysis}

GSA approaches rely on sampling based methods for uncertainty propagation, willing to fully map the
space of possible model predictions from the various model uncertain input parameters and then, allow
to rank the significance of the input parameter uncertainty contribution to the model output variability
\cite{Baroni14}. GSA approaches are well suited to be applied with models having nonlinear
behavior and when interaction among parameters occurs \cite{SaintGeours12}. These approaches going
through an intensive sampling are computationally demanding, as they most often rely on Monte-Carlo
(MC) approach, even though some more parsimonious sampling method such as Latin hypercube or
pseudo-Monte Carlo are sometimes applied (see \cite{Helton06} for a review). Most commonly,
GSA approaches rely on:
\begin{itemize}
 \item screening methods, such as Morris method \cite{Morris91};
 \item Sobol indices computation, that considers the output hyperspace ($x$) as a function ($Y(x)$)
 and performs a functional decomposition \cite{Iooss11,Iooss15} or a Fourier
 decomposition (FAST method) of the variance.
\end{itemize}

As fully detailed in \cite{Iooss15}, screening techniques ({\it e.g.} Morris method)
allow to classify uncertain input parameters in three categories: those that have negligible effect;
those that have linear effect; and those that have nonlinear effects or effects in interaction
with other input parameters. Sobol indices (or variance-based sensitivity indices) will explain
the share of the total variance in the space of output due to each uncertain input parameter
and/or input interaction. 
\newline

 \sloppy GSA has started to be applied in 1D hydraulic modelling in practical applications for hierarchical
ranking of uncertain input parameters \cite{Alliau15,Hall05,Jung15,Nguyen15,Pappenberger08}. As for 1D, applying a GSA
to flooding issues in 2D modelling requires method awareness among the community, practical tools
development and computational resources availability. Moreover an analysis on spatialization of input
uncertain parameters and on output variable is specifically needed in 2D \cite{SaintGeours11}.
Recently, GSA using a screening method has been implemented in 2D flood modelling application
\cite{Willis14} tackling ranking of uncertain input parameters using points and zonal approaches.
Computation of sensitivity maps such as maps of Sobol index is a promising outcome that has been
achieved for other types of water related issues \cite{Marrel11}.  \fussy

\subsection{Objectives of the study}\label{subsec:Objectives of the study}

To date, UA and SA have not yet been performed to specifically study uncertainty in 2D urban flood
simulations related to HR classified topographic data integration. Indeed, due to the curse of
dimensionality, SA methods have seldom been applied to environmental models with both spatially
distributed inputs and outputs \cite{SaintGeours14}. Such a problematic raises needs of
specific tools, computational resource and methods application. Among SA methods, a Global Sensitivity
Analysis (GSA) is implemented in this study. GSA approach is selected over LSA as 2D overland flow
process simulation through SWE system of partial differential equations, is viewed as being largely
nonlinear, with discontinuous solution and interactions between parameters. 
\newline

This paper aims to study uncertainty related to HR topographic data integration in 2D flood modelling
approach. The objective of the study is to perform an UA and a SA on two categories of uncertain
parameters (measurement errors and uncertainties related to operator choices) relative to the use of
HR classified topographic data in a 2D urban flood model having spatial inputs and outputs. Among SA
methods, a Global Sensitivity Analysis (GSA) is implemented to produce sensitivity maps based on Sobol
index computation. Carrying out these objectives will demonstrate the feasibility, the added values
and limitations of UA and SA implementation in 2D hydraulic modelling, in a context where spatial
variability and interaction are likely to occur. Moreover, modeller knowledge about challenges and
expectations related to HR classified data use in HR urban flood modelling will be enhanced. 
\newline

The study case is the low Var river valley (Nice, France) where flooding events occurred in the last
decades in the highly urbanized downstream part of the valley \cite{Guinot03}.
The output of interest is the overland flow water level ($Y(x)$). The used HR DEMs are based on
classified 3D dataset created from photointerpretation procedure. A proof of concept of GSA application
to 2D Hydraulic modelling voluntary choosing a resource requiring problem has been developed and
the method applied over an innovative concern related to the use of HR topographic data.
\newline

Following this introduction (Part \ref{sec:intro}) , the next part of the paper (Part \ref{sec:Method}) introduces the test case
context for SA methods uses, then enhances description of used HR topographic dataset, gives overview
of implemented methodology for the spatial GSA and introduces developed tools. The third part (\ref{sec:Results}) of
the paper presents results of UA and GSA, first at punctual then at spatial levels. The last
parts (Part \ref{sec:discussion} and \ref{sec:Summary-and-conclusions}) discuss outcomes and limits of our approach, providing concluding remarks.

\section{Method}\label{sec:Method}

The study area is a $17.8\;\text{km}^2$ domain that represents the last five downstream kilometers
of the low Var valley, located in Nice, France (figure \ref{fig1}). In the test basin, two major river flood
events occurred in last decades ($5^{th}$ of November $1994$; $6^{th}$ of November $2011$).
A HR topographic data gathering campaign fully covered the domain in $2010-2011$. The characteristics
of the river basin and of the $1994$ flood event are described in \cite{Guinot03}.
Between $1994$ and $2010-2011$ (date of event used for simulation and the date of the HR topographic
data gathering campaign), the studied area has considerably changed. Indeed, levees, dikes and urban
structures have been implemented, changing physical properties of the river/urban flood plain system.
Thus, the objective is not to reproduce the event, but simply to use the framework of this event as a
case study to carry out the UA and the SA. As mentioned in the introduction section, a GSA approach
using Sobol index is suitable to compute sensitivity maps \cite{Marrel11}. The method
implemented to carry out the spatial GSA is presented in detail in \cite{Abily15}, and
the upcoming description of the implemented GSA gives to the reader a summary of key elements
for understanding.

\subsection{HR classified topographic data and case study}\label{subsec:HR classified topographic data and case study}

\begin{figure}[htbp]
\begin{center}
\includegraphics[width=0.96\textwidth]{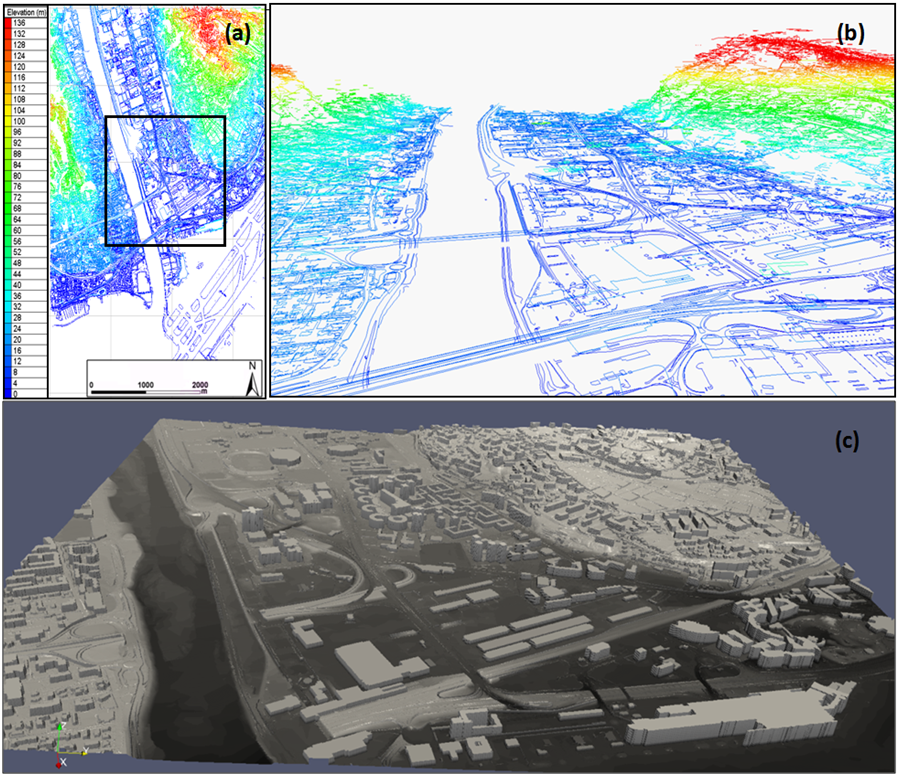}
\caption{Overview of the classes of photo-interpreted topographic data uses over the study area
(a, b) and HR DEM of a sub-part of interest of the domain (c).}
\label{fig1}
\end{center}
\end{figure}

The design and the quality of a photo-interpreted dataset are highly dependent on photogrammetry
dataset quality, on classes' definition, and on method used for digitalization of vectors
\cite{Lu07}. Reader can find details regarding principle of modern aerial photogrammetry
technology in \cite{Egels04}. The photogrammetric campaign carried out over the lower Var
valley, at a low flight elevation, allowed a pixel resolution at the ground level of $0.1\;\text{m}$ and had
a high overlapping ratio ($80\%$) among aerial pictures. Consequently, these characteristics allowed
the production of a high quality photogrammetric dataset. Using the photogrammetric dataset, the
photo-interpretation process has been carried out, to create a classified vectorial dataset through
digitalization of classes of polylines, polygons and points. This photo-interpreted dataset has been
designed with a total number of $50$ different classes representing large and thin above ground
features ({\it e.g.} buildings, concrete walls, road-gutters, stairs, {\it etc.}). The
specificities of the given photo-interpretation process can be found in \cite{Andres12}. For
hydraulic modelling purpose, $12$ classes over the $50$ are considered to represent above ground
features impacting overland flow path, as shown in figure \ref{fig1} (a, b). 
\newline

Classified data mean horizontal and vertical accuracy is $0.2\;\text{m}$. Errors in photo-interpretation which
results from feature misinterpretation, addition or omission are estimated to represent $5\%$
of the total number of elements. To control average level of accuracy and level of errors in
photo-interpretation, the municipality has carried out a terrestrial control of data accuracy
over $10\%$ of the domain covered by the photogrammetric campaign.

For our application, the 3D classified data of the low Var river valley is used to generate specific
DEM adapted to surface hydraulic modelling. For the GSA approach (see next section), only the input
DEM changes from one simulation to another and the hydraulic parameters of the model are set identically
for the simulations. Hydraulic conditions of the study case implemented in the models can be summarized
as follow: a constant discharge of $1,500\;\text{m}^3.\text{s}^{-1}$ is applied as input boundary
condition to reach a steady flow state condition almost completely filling the Var river bed.
This steady condition is the initial condition for the GSA and a $6$ hours estimated hydrograph
from the $1994$ flood event is simulated \cite{Guinot03}. The Manning's friction
coefficient ($n$) is spatially uniform on overland flow areas with a standard value of $0.015$, which
corresponds to a concrete surface. No energy losses properties have been included in the 2D hydraulic
model to represent the bridges, piers or weirs. Downstream boundary condition is an open sea
level (Neumann boundary condition) to let water flows out. 

\subsection{Spatial GSA approach}\label{subsec:Spatial GSA approach}

Falling within category of GSA approaches, screening methods allow in a computationally
parsimonious way, to discriminate among numerous uncertain input parameters those that have
little effect from those having linear, nonlinear or combined effects in output variance
\cite{Iooss15}. Screening methods principle consists in fixing an input parameters
set and performing an initial run. Then, for one parameter at a time, a new value of the parameter
is randomly chosen and a new run is performed. Variation in the run output is checked. This operation
is completed for all the parameters, $n$ times with $n$ equals to the total number of input
parameters. Screening methods perform well to discriminate influencing parameters on output
variability with 2D flood modelling studies \cite{Willis14}.
\newline
GSA approaches relying on Sobol index computation go one step further, allowing to quantify
the contribution to the output variance of the main effect of each input parameters
\cite{Sobol90,Saltelli99,SaintGeours12}. Sobol Index is based on functional
decomposition of variance (ANOVA), considering $Y$ the model output of interest as follow: $Y= f(X)$;
where $f$ is the model function, $X = (X_1;...;X_i)$ are $i$ independent input uncertain parameters
with known distribution. Sobol indices ($S_i$) of parameter $X_i$ are defined as:
\begin{equation}
 {S_i}_{(X_i)}=Var\left[ \mathbb{E}(Y|X_i)\right]/Var(Y),
\end{equation}
where $\mathbb{E}$ is the expectation operator. ${S_i}_{(X_i)}$ being the variance of conditional expectation
of $Y$ for $X_i$ over the total variance of $Y$, ${S_i}_{(X_i)}$ value will range between $\left[0; 1\right]$.
$S_i$ computations are computationally costly as it requires to explore the full space of inputs and
therefore an intensive sampling is necessary \cite{Iooss15}. 
\newline

Objective being to quantify impacts of input parameters, GSA approach using Sobol index is best suited
for sensitivity maps production. In this study, the implemented GSA follows standard steps used for
such type of approach as summarized in \cite{Baroni14} or in \cite{SaintGeours14}.
\newline

\begin{figure}[htbp]
\begin{center}
\includegraphics[width=0.98\textwidth]{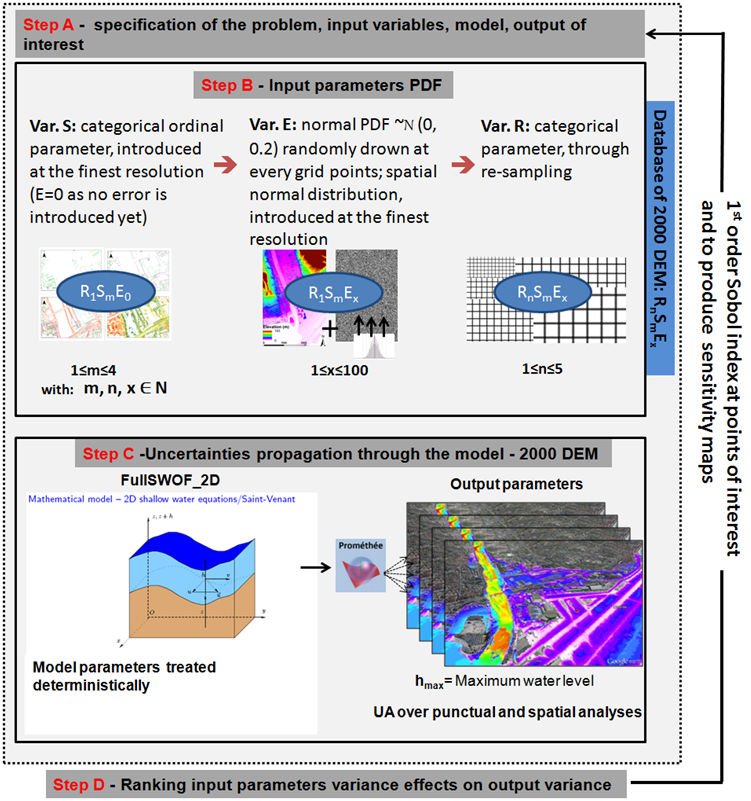}
\caption{Overview of the applied spatial GSA framework.}
\label{fig2}
\end{center}
\end{figure}

The steps of the method are presented in the figure \ref{fig2}: specification of the problem notably by
choosing uncertain parameters and output of interest (step A); assessing Probability Density
Function (PDF) of uncertain parameters (step B); propagating uncertainty, using a random sampling
approach in our case (step C); ranking the contribution of each input parameters regarding the output
variance (step D).
\newline

First steps of the approach (A and B) are the most subjective ones. For the study purpose, steps
A and B are treated as follow. Three input parameters related to uncertainties when willing to
use HR 3D classified data in 2D Hydraulic models are ($i$) one parameter related to the topographic
input error (called var. $E$) and ($ii$) two parameters related to modeller choices, when including
HR data in 2D hydraulic code (called var. $S$ and var. $R$) are considered in this GSA practical
case. These three parameters are considered as independent.
\newline

First, the uncertainties related to measurement errors in HR topographic dataset are considered
through var. $E$. This parameter is an error randomly introduced for every point of the highest
resolution DEM ($1\;\text{m}$) following a draw according to a normal distribution PDF, where the standard
deviation is equal to the RMSE value $\left(0.2\;\text{m}\right)$: $\mathcal{N} (0, 0.2)$. As from one point to the next one, the normal
PDF is drawn independently, it results in a spatialization that follows a uniform distribution.
Hundred maps of var. $E$ are generated and combined with the ($1\text{m}$) resolution DEMs.
\newline

Then, for uncertainties related to modeller choices when including HR data in hydraulic code, two variables
are considered: var. $S$ and var. $R$.
\newline

Var. $S$ is a categorical ordinal parameter having values representing the level of above ground
features details impacting flow direction included in DSM. $S1$ is a DTM (Digital Terrain Model)
only, $S2$ is $S1$ combined with buildings elevation inclusion, $S3$ is $S2$ completed with walls,
and $S4$ is $S3$ plus thin concrete structures (sidewalks, roar-curbs, {\it etc.}).
\newline

Var. $R$ represents choices made by the modeller concerning the computational grid cells resolution
in the model. In the hydraulic code used for this study (FullSWOF\_2D described in next
subsection), the grid cells are regular. This parameter var. $R$ can have five discrete values
from $1\;\text{m}$ to $5\;\text{m}$. At 1m resolution, number of computational points of the grid is above
$17.5$ million and at the $5\;\text{m}$ resolution grid size is $700,000$ computational points. The bounding
of this parameter is justified as on one hand, a grid resolution lower than $1\;\text{m}$ would result
in prohibitive computational time. On the other hand, resolutions coarser than $5\;\text{m}$ do not sound
to be a relevant choice for a modeller willing to create a HR model, as up-scaling effects would make the
use of the HR topographic data that are used as input irrelevant.
\newline

A total of $2,000$ DEMs are generated and used in the implementation of the GSA. The DEMs generation
process (step B, figure \ref{fig2}), explains as follow. Four DEMs at the finest resolution
(var. $R=1\;\text{m}$) are generated to implement all of the four var. $S$ possible scenarios. Then each
of these four DEMs is combined with the hundred var. $E$ grids producing $400$ DEMs. Eventually,
the $400$ DEMs combining all of the var. $S$ and var. $E$ possibilities combinations are
resampled to resolution $2,\;3,\;4,\;5\;\text{m}$ creating a database of $2,000$ DEMs where all the defined
input parameters can possibly be combined. 
\newline

\begin{figure}[htbp]
\begin{center}
\includegraphics[width=0.98\textwidth]{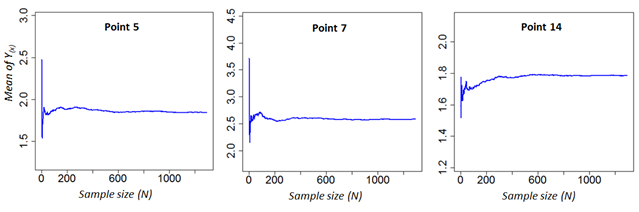}
\caption{Asymptotic convergence of random sampling at $3$ points of interest
(points $5$, $7$ and $14$ located on figure \ref{fig4}).}
\label{fig3}
\end{center}
\end{figure}

The propagation of uncertainty (step C) is carried out using a MC approach to randomly sample in
the database. Non-parsimonious approach consisting in computing a maximum number of simulations
among the $2,000$ possible cases has been carried out to generate a database of results. In total,
$1,500$ simulations out of the $2,000$ possible were computed to feed the result database using
the available $400,000$ CPU hours on a Cluster (cluster described in next sub-section). Therefore,
to make sure that the input space would be extensively explored, for all of the $20$ possible
var. $R$/ var. $S$ combinations, at least $50$ over the $100$ possible var. $E$ drawn were performed.
As the exploration of the space of input is restricted to $1,500$ simulations over $2,000$ possible cases,
an evaluation of the convergence is performed to assess if the convergence of the MC method is reached.
Figure \ref{fig3} illustrates the evolution of the convergence of the mean of the hyperspace of the
output of interest $Y(x)$ for three points (points located in figure \ref{fig4}), increasing $N$
through a random sampling in the result database. It is reminded here that the output of interest
is the simulated maximal overland flow water depth. An asymptotic convergence of the MC method is
observed for the three points, respectively when the sample size ($N$) is larger to $900$
simulations. Globally, over the 20 selected points, when N reaches a threshold value between
$900-1000$, the stabilization of the convergence is observed. 

Step D consists in the computation of Si using the output database. Sobol index of var. $R$, var. $S$
and var. $R$, respectively $S_i(R)$, $S_i(S)$ and $S_i(E)$ are computed following Eq. (1) at
points of interests. Spatialization of GSA approach is based on discrete realization of spatially
distributed input variables as described in \cite{Lilburne09}, and discrete computation
of output to produce sensitivity maps (as described in \cite{Marrel11}).

\subsection{Parametric environment and 2D SWE relying code}\label{subsec:Parametric environment and 2D SWE relying code}

Prom\'eth\'ee (a parametric modelling environment), has been coupled with FullSWOF\_2D (a 2D free
surface modelling code) over a High Performance-Computing (HPC) structure \cite{Abily15}.
\newline

Prom\'eth\'ee is an environment for parametric computation that allows carrying out uncertainties
propagation study, when coupled (or warped) to a code. This software is freely distributed by
IRSN (\url{http://promethee.irsn.org/doku.php}). Prom\'eth\'ee allows the parameterization with any numerical
code and is optimized for intensive computing. Moreover, statistical post-treatment, such as UA
and SA can be performed using Prom\'eth\'ee as it integrates R statistical computing environment \cite{Ihaka98}.
\newline

FullSWOF\_2D (for Full Shallow Water equation for Overland Flow in 2 dimensions) is a code
developed as free software based on 2D SWE \cite{Delestre12,Delestre14b}. Two parallel versions
of the code have been developed allowing to run calculations under HPC structures
\cite{Cordier13}. In FullSWOF\_2D, the 2D SWE are solved using a well-balanced finite
volume scheme based on the hydrostatic reconstruction (see \cite{Audusse04c,Delestre14b}).
The finite volume scheme is applied on a structured spatial discretization using regular
Cartesian meshing. For the temporal discretization, based on the CFL criterion, a variable time step
is used. The hydrostatic reconstruction (which is a well-balanced numerical strategy) allows
to ensure that the numerical treatment of the system preserves water depth positivity and does
not create numerical oscillation in case of a steady states, where pressures in the flux are balanced
with the source term here (topography). Different solvers can be used HLL, Rusanov, Kinetic
\cite{Bouchut04}, VFRoe-ncv combined with first order or second order (MUSCL or ENO) reconstruction.
The HLL solver has been used in this study with a first order MUSCL reconstruction method. 
\newline

On the HPC structure (Interactive Computation Centre of Nice Sophia Antipolis University), up
to $1152$ CPUs are available and up to $30$ simulations can be launched simultaneously using
Prom\'eth\'ee-FulSWOF\_2D. A database of flood maps results has been produced using a total of
$400,000$ CPU hours. The required unitary computation time is two hours over $64$ CPUs, for simulations
using the finest resolution grid size ($1\;\text{m}$), which has $17.8$ millions of computational points. At
the coarsest resolution ($5\;\text{m}$), the grids size is decreased to $712,000$ computational points and using
$64$ CPUs, the computational time decreases to few minutes.

\section{Results}\label{sec:Results}

This section presents the results of the UA and the GSA. A subarea is selected in the flooded
area of the domain to carry out the spatial analysis. This subarea is $4.35\;\text{km}^2$,
representing one quarter of the total spatial extent of the model. $20$ points of interest are
defined in the selected flooded area of the subarea (figure \ref{fig4}). Points $1$ to $10$ are
spread in and around the main streets. These streets are densely urbanized. Points $11$ to $16$
are located in less urbanized areas (stadium, parking, small agricultural field, {\it etc.}).
Moreover, from points $15$ to $20$, points are located in areas which are at the edge of the flood
extent, either in open area (points $15$ and $16$) or where above ground features are densely
present (points $17$ to $20$).

\subsection{Uncertainty analysis}\label{subsec:Uncertainty analysis}

\subsubsection*{Punctual view}

\begin{figure}[htbp]
\begin{center}
\includegraphics[width=0.98\textwidth]{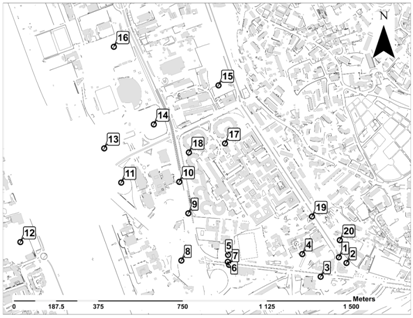}
\caption{Location of points of interest and associated values.}
\label{fig4}
\end{center}
\end{figure}

Mean and variance of computed maximal water depth ($Y(x)$) at the different points of interest
are presented in figure \ref{fig4}. Means and standard deviations of $Y(x)$ values are computed
using the full size database ($N=1,500$). Over the $20$ points of interest, importance of the variability
introduced by uncertain input parameter is significant ($0.51\;\text{m}$ in average). Moreover, variability
in $Y(x)$ variance can be important as the minimal variance is $0.28\;\text{m}$ (point $17$), and the maximal
variance is $0.71\;\text{m}$ (point $8$). Further interpretation, such as the analysis of the trend in the
magnitude of variance changes from one point to another, is not accessible for generalization
using punctual observation only. Nevertheless, studying the distribution of $Y(x)$ at points of
interest gives another insight to carry out the uncertainty analysis.

\begin{figure}[htbp]
\begin{center}
\includegraphics[width=0.98\textwidth]{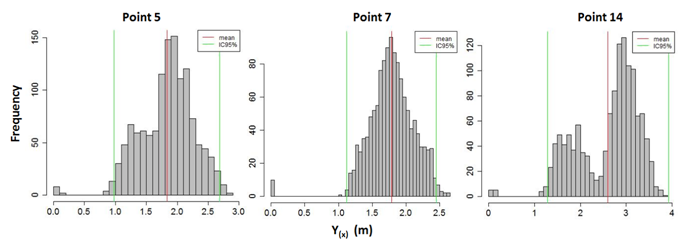}
\caption{Distribution of $Y(x)$ at three points of interest (points, located on figure \ref{fig4}).}
\label{fig5}
\end{center}
\end{figure}

Figure \ref{fig5} illustrates $Y(x)$ distributions using the complete set of available model runs in the database for
three points. $Y(x)$ follows a normal distribution, as observed for point $7$ or distribution can be bi-modal
as observed for point $14$. The difference between the normal and bi-modal distribution of $Y(x)$ is not always
clearly observed (point $5$). Most of the clearly observed bimodal distribution (ten out of the twenty points)
occurs for the points located in the central part of the highly urbanized area (points $1$ to $10$). This area is
largely flooded, and seven points have here a clearly marked bimodal distribution. In largely flooded but
relatively less urbanized areas, the trend is reversed as five out of six points have a normal distribution.
Lastly for the points located at the edge of the flooded areas, two over four points have a bimodal distribution,
whereas the two others have a normal one. Bimodal distributions lead to larger amplitude in $Y(x)$ distribution.
The bimodal distribution illustrates the non–linearity between the input and output. Explanations to link these
observations with physical properties of phenomena and of uncertain input parameter properties are given,
combining these observations with SA results, in the discussion section. Moreover, it is noticeable that points
are sometimes not flooded at all, when $Y(x)$ is equal to zero. Reasons for these zero values are that in seldom
cases, var. $E$ value gives at the point of interest a high ground elevation value (above $Y(x)$ value) or that
var. $S$ produces critical threshold effects diverting flow direction. 

\subsubsection*{Spatial analysis}

\begin{figure}[htbp]
\begin{center}
\includegraphics[width=0.98\textwidth]{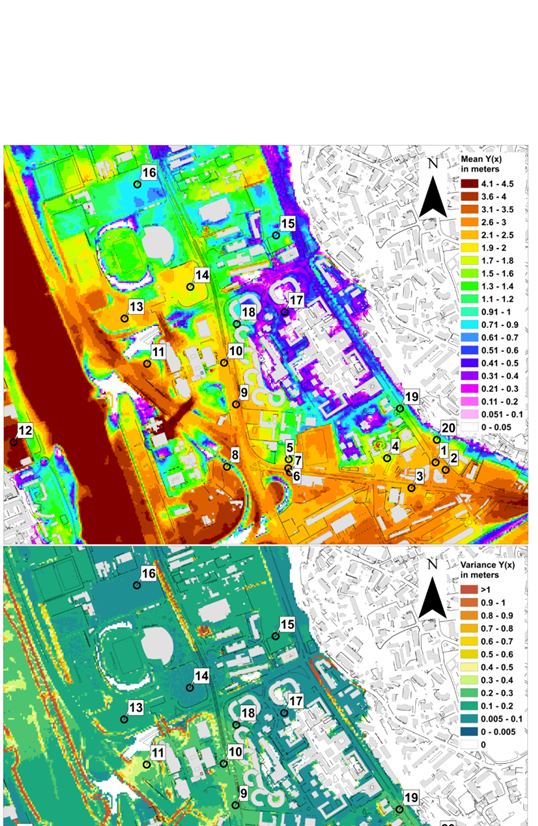}
\caption{Maps of mean and variance values of $Y(x)$.}
\label{fig6}
\end{center}
\end{figure}

The comparison of maps of $Y(x)$ mean and variance (figure \ref{fig6}) puts to the light the fact that areas
densely urbanized and having a high water depths, have a high variance in $Y(x)$. This maps comparison also
underlines the fact that areas having a high mean water depths in less densely urbanized, and in areas close to
the edge of flood spatial extent (having a smaller mean water depth) have a lower variance value of $Y(x)$.
This confirms the local observations at points of interests. Moreover, a high variance of $Y(x)$ is observed in
the map for places that have steep slope such as river bank, access roads, highway ramps or dikes. Intuition
would lead to incriminate here resolution of discretization effects (var. $R$) as it will be confirmed by
the SA (see next section and discussion part). Over the river, variance is locally important. The spatial changes
in variance in the river bed ranges from $0.1\:\text{m}$ to $1\;\text{m}$. Amplitudes in variance in the river bed are most likely
due to above ground features additions when var. $S$ changes (features such as walls, dikes levees, and roads
elements in the main riverbed), that does change the width of the river bed itself. Consequently, these local
important variance values are not surprising. Our study focuses on overland flow areas. A GSA over riverbed itself
would be out of the range of the spatial GSA defined for this study.

\subsection{Variance based global sensitivity analysis}\label{subsec:Variance-based-global}

\subsubsection*{Flood event scenario}

\begin{figure}[htbp]
\begin{center}
\includegraphics[width=0.98\textwidth]{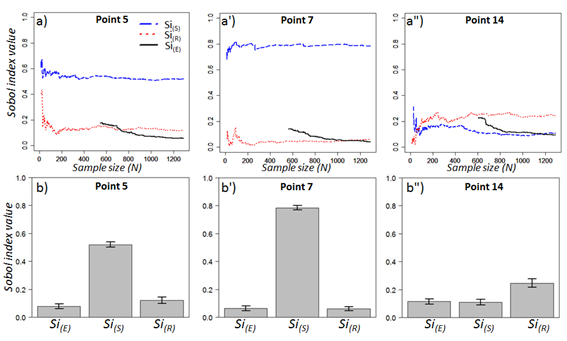}
\caption{Illustration for three points of interest of Sobol indices convergence (a, a', a'') and of confidence interval
 computed using bootstrap method (b, b', b'').}
\label{fig7}
\end{center}
\end{figure}

$1^\text{st}$ order Sobol index ($S_i$) of var. $S$ ($S_i(S)$), var. $R$ ($S_i(R)$) and var. $E$ ($S_i(E)$) are
computed for the $20$ points of interest. Figure \ref{fig7} (a) shows the evolution of computed $S_i$ increasing
$N$ through a random sampling in the results database for the same three points used in the figure \ref{fig5}.
Stabilization of the computed $S_i$ values is observed when $N$ is approximately $1,000$, confirming that
convergence of the random sampling is reached around this $N$ value. It has to be noticed that below a value
of $N=500~600$, the samples are too small to compute $S_i(E)$ with our algorithm (draws of var. $E$ are too
scarcely distributed in the matrix to compute conditional expectation of var. $E$). A bootstrap is performed,
to check confidence interval of the computed $S_i$ as can be seen in figure \ref{fig7} (b). For each point,
independent samples of size $N=1,000$ are randomly drawn $10,000$ times in the results data base to compute
$10,000$ times $S_i$. Then the $S_i\;95\%$ confidence interval is computed.\\

Over the $20$ selected points, the average $S_i(S)$ value is $0.40$, the average $S_i(R)$ value is $0.24$ and the
average $S_i(E)$ value is $0.06$. $S_i(S)$ is ranked as the highest among the three $S_i$ for $13$ out of the $20$
points. For the seven other points, $S_i(R)$ is ranked as the highest $S_i$. The results show that Var. $E$ is
never the variable which influences the most $Y(x)$ variance and $S_i(E)$ is ranked as the second highest $S_i$
only for points $15$ and $16$. These points are located at the edge of the flood extent area where the $Y(x)$ values
are in average below $1\;\text{m}$.\\

For the $20$ points, the difference between the highest ranked $S_i$ and the second one is often clear (around $0.35$),
but the difference between the $S_i$ ranked as $2^\text{nd}$ and $3^\text{rd}$ is often not important (around $0.1$) and can
be smaller than the $95\%$ confidence interval calculated from the bootstrap.\\

The main outcome from the punctual GSA is that var. $S$ and var. $R$, which are two modeller choices when including
HR topographic data in the model, are always the parameters contributing the most to $Y(x)$ variance. The analysis
also highlights that var. $E$ does not introduce much variance on $Y(x)$. For the $20$ points, $S_i$ ranking varies
from one point to another one, enhancing the spatial variability of uncertain parameters influence on $Y(x)$
variance and strengthening the interest of sensitivity maps production.

\subsubsection*{Spatial analysis}

Over the selected subarea, $S_i$ are computed every $5\;\text{m}$ to produce sensitivity maps. With this level of
discretization, it represents a total of $120,000$ points where $S_i$ are calculated. A test has been carried out
at a finer resolution ($1\;\text{m}$) over a $100\;\text{m}$ per $100\;\text{m}$ area for $S_i$ mapping. Results in $S_i$ maps
at $1\;\text{m}$ and $5\;\text{m}$ are similar over this small area. Therefore, the $S_i$ maps are computed at a resolution
 $5\;\text{m}$ as the number of points to compute is $25$ times less important than for a $1\;\text{m}$ resolution.
 The $S_i$ are computed at every points
using $N$ equal to the full size of available simulations in the results database ($1,500$).\\

\begin{figure}[htbp]
\begin{center}
\includegraphics[width=0.98\textwidth]{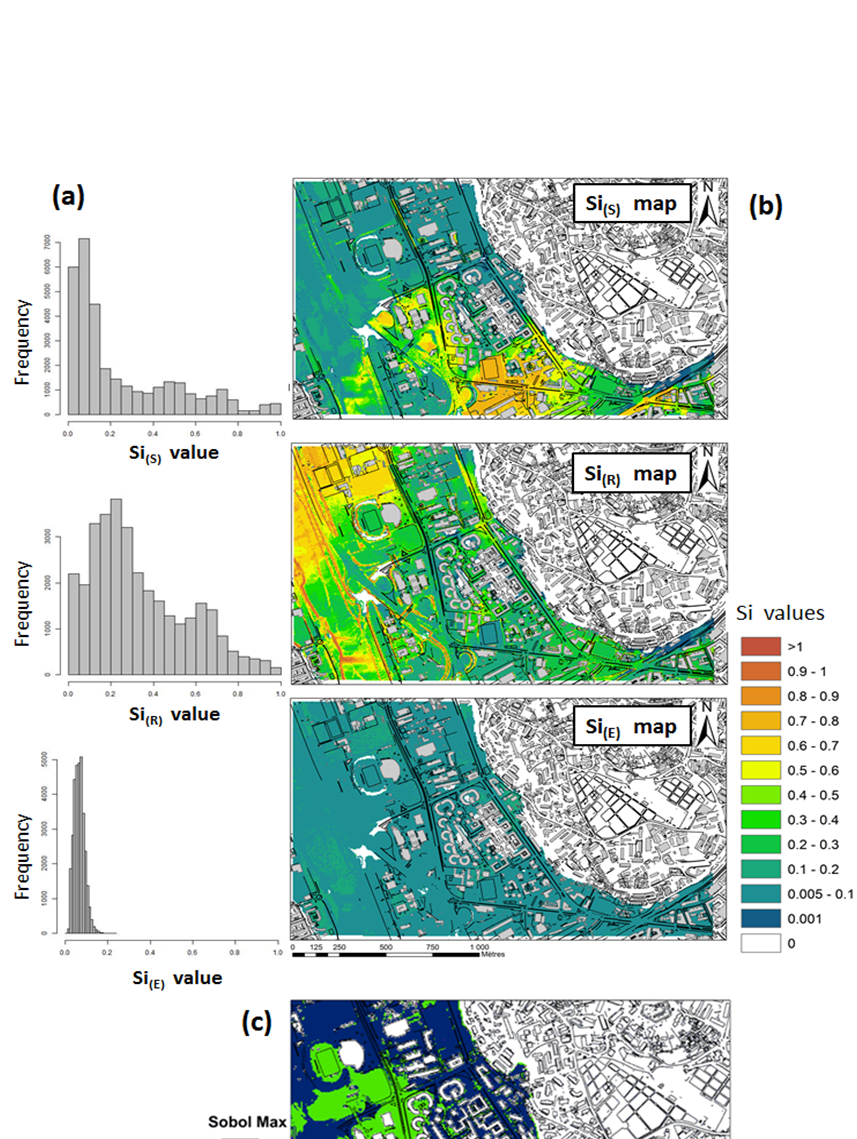}
\caption{Distribution of computed $S_i$ (a), details of $S_i$ maps (b) and map of highest ranked $S_i$ (c).}
\label{fig8}
\end{center}
\end{figure}

A first analysis of the distribution of computed $S_i$ is illustrated in figure \ref{fig8} (a). Non flooded areas
are removed for this analysis as well as areas covered by buildings. Indeed, inside the buildings which are
represented as impervious blocks in the model, $S_i(S)$ is equal to one. Therefore, var. $S$ explains the
entire variance of $Y(x)$ in building areas. Moreover, at the edges of buildings, $S_i(R)$ is equal to one as
well, due to buildings resolution effects. The number of points where $S_i$ have been calculated and that are
plotted in figure \ref{fig8} (a) is around $60,000$. The results show that:

\begin{itemize}
 \item $S_i(S)$ is highly distributed around $0.1$ and has two peaks in distribution around $0.6$ and $0.75$ that have
 a flatter shape;
 \item $S_i(R)$ is highly distributed around a value of $0.25$. A second minor distribution peak around $0.60$;
 $S_i(E)$ distribution is a single peak centered in $S_i(E)=0.07$, which is a value lower than both $S_i(R)$
 and $S_i(S)$ peaks.
\end{itemize}

Analysis of these multi-modal distributions, confirms punctual GSA results regarding the non-spatially homogeneous
ranking of $S_i$. The analysis of $S_i$ maps will help to understand the spatial distribution and the ranking of
uncertain input parameters according to their influence over the output variance.\\

Figure \ref{fig8} (c) presents the Sobol index maps. Analyzing in the first place the maximal $S_i$ spatial
distribution, it appears that, $S_i(R)$ and $S_i(S)$ are always ranked with the highest value. $S_i(R)$ is ranked
as the highest over $67\%$ of the subarea whereas $S_i(S)$ is ranked as the highest over $32\%$ of the subarea. Var. $E$
is rarely the most impacting parameter. This confirms the punctual GSA results and the $S_i$
distribution analysis. In the second place, using the spatial repartition of $S_i$ values presented as sensitivity
maps (figure \ref{fig8} (b)), the following remarks arise:
\begin{itemize}
 \item $S_i(s)$ is ranked as the highest index in locations where $Y(x)$ has a high variance. In the areas with
 a high $Y(x)$ variance, $S_i(s)$ values range between $0.3$ and $0.8$. Those high $S_i(s)$ areas are characterized by
 a highly urbanized environment where above ground features strongly impact $Y(x)$.
 \item $S_i(S)$ is ranked as the highest $S_i$, where a given above ground element strongly impact locally
 hydrodynamic and consequently $Y(x)$.
 \item $S_i(R)$ happens to be the most impacting parameters in areas less densely urbanized.
 \item Moreover, high ranking of $S_i(s)$ also occurs when a given aboveground structure impacts upstream or
 downstream calculation of $Y(x)$ whatever is the urban configuration/density of affected upstream or downstream
 areas.
 \item $S_i(R)$ is ranked as the highest $S_i$ when $Y(x)$ is low (below $1\;\text{m}$), and when in the meantime, variance
 of $Y(x)$ is low as well. It corresponds to areas close to the edge of the flood extent.
 \item $S_i(R)$ is ranked as the highest $S_i$ in areas which are less densely urbanized and where no above ground
 features, at the given area, neither upstream nor downstream, have any important effects on $Y(x)$.
 \item $S_i(R)$ is ranked as the highest $S_i$ in areas where the ground slope is steep. Indeed the level
 representation of a sloping area is highly affected locally by the degree of resolution of the discretization.
 \item $S_i(E)$ low and almost homogeneous over the subarea.
\end{itemize}

\section{Discussion}\label{sec:discussion}

The Implemented approach is a proof of concept of applicability of spatially distributed GSA to 2D hydraulic
problems. UA and spatial ranking of influent uncertain input parameters over the 2D HR flood modelling
study case have been achieved. Nevertheless, being a first attempt, the approach can be improved. Outcomes,
limits and perspectives are underlined in this section and compared with other research fields in geomatics,
SA and hydraulic modelling.

\subsection{Outcomes}\label{subsec:Outcomes}

A basic UA leads to the following conclusions on: output variability quantification, nonlinear behavior of the
model and spatial heterogeneity. Within established framework for the UA, the considered uncertain parameters
related to the HR topographic data accuracy and to the inclusion in hydraulic models influence the variability
of $Y(x)$ in a range that can be up to $0.71\;\text{m}$. This stresses out the point that even though hydraulic parameters
were set-up as constant, the uncertainty related to HR topographic data use cannot be omitted and needs to be
assessed and understood. These warnings were already raised up in \cite{Dottori13} and \cite{Tsubaki13},
and are strengthened in this study by $Y(x)$ variance quantification. The quantification is not
easily transposable in other contexts and it is not an easy process to give general trend for practical
applications given the fact that ($i$) spatial heterogeneity of $Y(x)$ variance is observed and ($ii$)
specificities of different HR classified dataset is highly variable. Nevertheless, this quantification of
uncertainty goes in the direction of improvement of state of the art as common practice is still to quantify
uncertainty using expert opinion only (see \cite{Krueger12}). Investigations on the UA can lead to deeper
understanding of mechanisms leading to $Y(x)$ variability. The punctual analyses of the $Y(x)$ distributions
(either unimodal or multimodal) illustrate the nonlinearity of uncertain parameters effects over the output.
This nonlinearity in the output distributions is most likely due to var. $S$ which represents the level of
details of above ground features incorporated in HR DEMs.\\

Punctual SA highlighted that depending on location of considered point of interest, maximal first order $S_i$ are
different. This goes in the direction of a need of spatial representation of $S_i$ under the form of sensitivity
maps for consistent analyses. This spatial distribution of $S_i$ showed the major influence of the modeller
choices when using the HR topographic data in 2D hydraulic models (var. $S$ and var. $R$) with respect to
the influence of HR dataset accuracy (var. $E$). Hence as underlined in \cite{Marrel11}, if one wants to
reduce variability of $Y(x)$ at a given point of interest, the use of sensitivity maps helps to determine the
most influential input at this point. Moreover, sensitivity maps give possibility to link the spatial distribution
of $S_i$ to the properties of the model, especially with the physical properties of represented urban sector
topography. The fact that var. $S$ is the most contributing parameter in densely urbanized areas is not
surprising as it introduces a change in the representation in the model of physical properties of the urban
environment. The var. $R$ indirectly impacts quality of small scale elements representation well. 

\subsection{Limits of the implemented spatial GSA approach}\label{subsec:Limits-implemented-GSA}

GSA allowed to compute sensitivity maps, but simplifications and choices, especially regarding the way step
A (setting up of the spatial GSA framework by choosing uncertain parameters and choosing a way to spatialize
them) and step B (assigning PDF to input parameters), lead to simplifications which are interesting to enhance.\\

For the uncertainties related to errors in HR topographic data (var. $E$), the followed Normal PDF having properties
of the RMSE is randomly introduced, for every points of the highest resolution DEM ($1\;\text{m}$). Nevertheless, as from
one point to the next one, the normal PDF is drawn independently, it results in a uniform spatial distribution.
In practice a uniform repartition should increase entropy and maximize errors/uncertainties effects. In the present
case, this consideration is not valid. Indeed, the used parameter is a RMSE which is already averaged over the
space. In fact as reminded in \cite{Wechsler07}, the RMSE is calculated based on assumption of normality
which is often violated. For instance, over open and flat areas ({\it e.g.} parking, roads), relative accuracy
from one point to another should increase. In the study case, a comparison with ground topographic data
measurement revealed that accuracy of HR DEM RMSE increases to $0.05\;\text{m}$. Hence, over flat areas where
the var. $E$ appears to be ranked as the second most contributing parameter to $Y(x)$ variability, not
without standing the fact that the $S_i$ confidence interval of ranked second and third parameters overlaps,
it sounds reasonable to think that var. $E$ is overestimated. Opposite effect is observable over sloping
areas ({\it e.g.} dikes), where in our cases, after a regional control of the measurement quality, it is found
that RMSE value is about $1\;\text{m}$. Therefore, especially over steep slope areas such as dikes where var. $R$ has been
found to be the most important parameter contributing to $Y(x)$ variability, our $S_i$ ranking has to
be taken with caution as var. $E$ has probably been locally underestimated. For further work, it would be
interesting to improve the approach, by using spatialized value of RMSE in function of topographic properties.
This regionalization of characteristics of PDF might not be easy to implement by practitioners as regional
information of accuracy might not be available. In that case, basic assumption to attribute regionally
different characteristics to PDF could be relevant. For var. $E$, a component related to photointerpretation
errors should have been taken into consideration. Moreover, in order to improve our study, it would be
relevant to include a new variable that would reflect errors in photointerpretation. Basically, this should
consist in a random error in classified data for $5\%$ of the number of elements used for DEM generation.
From a technically point of view, implementation of such process is not straight forward particularly,
recalling that this study is a first proof of concept on the topic. Therefore, it has not been included in
the SA. Nevertheless, errors in photo-interpretation, which are uncertainties inherent to the HR dataset would
have locally a considerable impact on $Y(x)$ variability and would require further research.\\

For modeller choices, in terms of level of details in classified features to be integrated in the hydraulic
models (var. $S$), it is reasonable to consider this parameter as a categorical ordinal parameter having a
uniform PDF. Indeed, depending on availability of information of features influencing overland flow defined as
classes and depending on model objective, modeller will select one of the available options in increasing
complexity of DEM. The choice of a row HR DTM (without buildings, var. $S1$) is mostly responsible of the
observed binomial distribution in the UA, leading to an under estimation of maximal water depth $Y(x)$ compare
to other cases. Nevertheless it appears as well that punctually, at $1\;\text{m}$ and $3\;\text{m}$ resolutions, var. $S4$ leads
to low $Y(x)$ value as well due to local effects over flow paths.\\

For modeller choices in terms of level of discretization (var. $R$), HR DEM were used, we constrained ourselves
to resolution levels which are realistic with the use of such type of data considering that a resolution higher
than $5\;\text{m}$ is not compatible with the idea of producing HR models. Nonlinear effects of resolution are long
time known by practitioners in the sense that the grid resolution will impact the level of details included
in the model \cite{Horritt01,Mark04,Djordjevic13}.

\section{Summary and conclusions}\label{sec:Summary-and-conclusions}

Implemented approach is a proof of concept of applicability of spatially distributed variance-based Global
Sensitivity Analysis (GSA) to 2D flood modelling, allowing to quantify and to rank the defined uncertainties
sources related to topography measurement errors and to operator choices when including High Resolution (HR)
classified dataset in hydraulic models. Interest focuses on ($i$) applying an Uncertainty Analysis (UA)
and spatial GSA approaches in a 2D HR flood model having spatial inputs and outputs and ($ii$) producing
sensitivity maps. Summary of outcomes and remarks are put to the front concerning these aspects.

\subsubsection*{Spatial GSA implementation}

\begin{itemize}
 \item Using $400,000$ CPU hours on the HPC architecture of the Centre de Calcul Interactif, a database of $1,500$
 simulations of a river flood event scenario over a densely urbanized area described based on a HR classified
 topographic dataset has been built. A random sampling on the produced result database was performed to follow
 a Monte-Carlo approach. After convergence check, a UA and a variance based functional decomposition GSA have
 been performed over the output of interest. Output of interest being the maximal overland flow water depth
 ($Y(x)$) reached at every point of the computational grids.
 \item Feasibility of spatial GSA approach for HR 2D flood modelling was achieved by this proof of concept
 test study.
 \item Important requirements are involved when implementing UA and GSA as expertise and efforts are required
 ($i$) for method establishment (specification of the problem) and ($ii$) for characterization of input parameters
 as complexity of this step increases to consider spatial variability of the input parameters and can involve
 an important pretreatment phase ({\it e.g.} for DEMs generation). Eventually spatial information of HR
 topographic dataset accuracy might not be available. In that case, basic assumption attributing regionally
 different characteristics to PDF could be relevant. Not only this part of the process is subject to
 subjectivity, but it can be time consuming and his application in dedicated tools (such as
 Prom\'eth\'ee-FullSWOF\_2D) might not be straight forward.
 \item For practical application, restrictive computational resources requirement is raised for this specific
 case (in terms of CPU and in terms of hard drive storage) due to the use of big data combined with a Monte
 Carlo approach. More parsimonious strategies like Pseudo Monte Carlo sampling could be used or, depending on
 objective other GSA method than used Sobol functional variance decomposition can be carried out: see \cite{Iooss15}
 for a review on optimization of GSA strategy in function of objectives and complexity of models.
 \end{itemize}

\subsubsection*{Uncertainties related to HR classified topographic data use}

\begin{itemize}
 \item The UA has allowed to quantify uncertain parameters impacts on output variability and to describe the
 spatial pattern of this variability. The spatial GSA has allowed the production of Sobol index ($S_i$) maps
 over the area of interest, enhancing the relative weight of each uncertain parameter on the variability of
 calculated overland flow. 
 \item Within established framework, the considered uncertain parameters related to the HR topographic data
 accuracy and to the inclusion of HR topographic data in hydraulic models influence the variability of $Y(x)$,
 in a range that can be up to 1 m. This enhances the fact that the uncertainty related to HR topographic data
 use is considerable and deserves to be assessed and understood before qualifying a 2D flood model of being
 HR or of high accuracy. Moreover, UA reveals non linear effects and spatial heterogeneity of $Y(x)$
 variance. Nonlinearity in the output distributions is most likely due to var. $S$ which represents the
 level of details of above ground features incorporated in DEMs.
 \item Quantification of uncertainty through UA goes in the direction of improvement of state of the art,
 compared to quantification of uncertainty based on expert opinion only. Investigations on the UA can lead
 to deeper understanding of mechanisms leading to $Y(x)$ variability. Moreover, an analysis of $Y(x)$ extreme
 quantiles distribution could have been performed to find the combination of penalizing parameters.
 \item The spatial distribution of $S_i$ illustrates the major influence of the modeller choices, when using
 the HR topographic data in 2D hydraulic models (var. $S$ and var. $R$) with respect to the influence of HR
 dataset accuracy (var. $E$). As underlined in \cite{Marrel11}, if one wants to reduce variability of $Y(x)$
 at a given point of interest, the use of sensitivity maps helps to determine the most influential input at
 this point. Moreover, possibility to link the spatial distribution of the $S_i$ to the properties of the
 model, especially with the physical properties of represented urban sector topography. The fact that
 var. $S$ is the most contributing parameter in densely urbanized areas is not surprising. Indeed, in that
 case, a change in var. $S$ highly influences the representation in the model of physical properties of
 the urban environment, therefore impacting model results. Var. $R$ indirectly impacts quality of small
 scale elements representation as well. Nevertheless var. $E$ assumes a spatially uniform RMSE and does not
 take into consideration errors in photo-interpretation. Therefore, errors related to HR measurement are
 probably underestimated locally in this study.
 \item GSA use to spatially rank uncertain parameters effects gives a valuable insight to modeller.
 Moreover, it can help to reduce variability in the output putting effort on improving knowledge about a
 given parameter or helps for optimization ({\it e.g.} to define relevant areas where spatial discretization
 is important prior to non structured mesh use).
 \item Quantification and ranking helps modeller to have a better knowledge of limits of what has been
 modelled. Nevertheless, as reminded in \cite{Pappenberger08} depending in the method GSA might produce
 different results.
\end{itemize}

\section*{Acknowledgements}\label{sec:Acknowledgements}

Photogrammetric and photo-interpreted dataset used for this study have been kindly provided by Nice C\^ote
d’Azur Metropolis for research purpose. The Nice C\^ote d’Azur Metropolis direction of geographic
information, particularly G. Tacet, F. Largeron and L. Andres gave comments of great value regarding Geomatic
aspects. Authors are thankful to CEMRACS 2013 organizers, Y. Richet and B. Ioss, which provided valuable
comments for GSA aspects. This work was granted access to the HPC and visualization resources of the
"Centre de Calcul Interactif” hosted by University Nice Sophia Antipolis.


\end{document}